# New Experimental Limit on Photon Hidden-Sector Paraphoton Mixing

A. Afanasev,[1]  O.K. Baker,[2]  K.B. Beard,[3]  G. Biallas,[4]  J. Boyce,[4]  M. Minarni,[5]  R. Ramdon,[1]  M. Shinn,[4]  and P. Slocum[2]

[1.] Department of Physics, Hampton University, Hampton, VA 23668
[2.] Department of Physics, Yale University, P.O. Box 208120, New Haven, CT 06520
[3.] Muons, Inc., 552 N. Batavia Avenue, Batavia, IL 60510
[4.] Free Electron Laser Division, Jefferson Laboratory, 12000 Jefferson Avenue, Newport News, VA 23606
[5] Department of Physics, Universitas Riau (UNRI), Pekanbaru, Riau 28293 Indonesia

We report on the first results of a search for optical-wavelength photons mixing with hypothetical hidden-sector paraphotons in the mass range between $10^{-5}$ and $10^{-2}$ electron volts for a mixing parameter greater than $10^{-7}$. This was a generation-regeneration experiment using the "light shining through a wall" technique in which regenerated photons are searched for downstream of an optical barrier that separates it from an upstream generation region.  The new limits presented here are approximately three times more sensitive to this mixing than the best previous measurement. The present results indicate no evidence for photon-paraphoton mixing for the range of parameters investigated.

The Standard Model (SM) of particle physics [1-5] provides a wonderfully successful, well-tested description of the strong, electromagnetic, and weak interactions between half-integer spin fermions and integer spin bosons at the smallest length scales and highest energies accessible in current experiments.  However it has limitations:  the apparent failure to explain dark energy and dark matter, an unnaturally small CP-violating parameter associated with the strong interaction, and 19 free parameters, to name a few.  If the SM is part of a more fundamental theory which has some new mass scale, new dynamics and particles would appear and hence signal the new physics associated with it. Popular extensions of the SM based upon string theory for example, predict a "hidden sector" of particles that interact with the "visible sector" SM fields only with feeble, gravitational-strength couplings [6-7]. This hidden sector can be probed using very high energy accelerators such as the Large Hadron Collider at the TeV scale, and also by laser experiments at the sub-electron volt (sub-eV) energy scale [8-20].

In this hypothesis, low energy dynamics involves the familiar massless electromagnetic force mediator photon, and additionally a hidden sector paraphoton which may have a finite mass.

The most general renormalizable Lagrangian describing the interaction dynamics of these two fields at low energies is [6]

$$L = -\frac{1}{4}F^{\mu\nu}F_{\mu\nu} - \frac{1}{4}B^{\mu\nu}B_{\mu\nu} - \frac{1}{2}\chi F^{\mu\nu}B_{\mu\nu} + \frac{1}{2}m_{\gamma'}^2 B_\mu B^\mu. \qquad (1)$$

Here $F_{\mu\nu}$ is the ordinary electromagnetic gauge field strength tensor, $B^{\mu\nu}$ is the field strength tensor for the hidden sector field $B^\mu$ and $m_{\gamma'}$ denotes the hidden sector paraphoton mass. The first two terms in (1) are the kinetic terms for the SM photon and hidden sector photon fields, respectively. The third term corresponds to a non-diagonal kinetic term, that is, kinetic mixing between the two fields. The last term of the Lagrangian indicates a possible mass for the paraphoton. The mixing parameter $\chi$ is predicted to range between $10^{-16}$ and $10^{-4}$ in some string theory based calculations [6-7]. However it is a completely arbitrary parameter and even $\chi = 0$ is possible. New limits are placed on this parameter in the work described here.

The importance of this study goes beyond even particle physics. A recent suggestion that paraphotons may give rise to a hidden cosmic microwave background (HCMB) [21] indicates that sub-eV particle physics may have direct bearing on cosmological studies. If there is photon-paraphoton resonant kinetic mixing, then a measurement of this mixing may provide new constraints on the effective number of neutrinos produced after nucleosynthesis and before CMB decoupling [21].

During the past couple of years, several experimental groups have obtained new data that may illuminate the hidden sector with its potentially small mixing with SM fields in the sub-eV energy range: GammeV [22], BMV [23], OSQAR [24], and PVLAS [25]. These first three experiments are all based upon the "light shining through a wall" technique [12, 26] where laser light impinges upon a wall that it cannot penetrate, and a search is made for photons that reappear behind the wall. Only the weakly interacting, small mass, new particle would penetrate the wall and give rise to a regenerated photon signal. Vacuum oscillations of photons ($\gamma$) into hidden-sector paraphotons ($\gamma'$) with sub-eV mass may yield nonvanishing regeneration rates in a carefully designed experiment if such particles exist [6]. The process is depicted in Fig. 1.

The Light Pseudoscalar and Scalar Search (LIPSS) collaboration took data that tests the $\gamma$-$\gamma'$ mixing in a series of runs at the Jefferson Lab (JLab) Free Electron Laser (FEL) facility in Spring 2007. The experimental setup is shown in Fig. 2 and is described in more detail, along with the experimental procedure, in [27]. A description is given here that is relevant for the hidden sector photon physics experimental study.

The FEL provided laser light for the LIPSS Experiment that was tuned to a wavelength of 0.935 ± 0.010 microns in pulses that were 150 fs long with a variable repetition rate of up to 75 MHz. The average intensity, monitored continuously during the experiment, was 180 watts.

The FEL beam passed through an optical transport system, was collimated to 8 mm beam diameter and was directed onto the LIPSS beam line through a series of water-cooled turning mirrors (TM's) and collimators, as shown in Fig. 2. The LIPSS beam line consisted of an upstream (generation) region and a separate (regeneration) region downstream of it. The generation region was three meters long; the downstream regeneration region was identical to it. Between the generation and regeneration regions was an optical beam dump that also served as a power meter. Turning mirror TM3 and the beam dump in combination with a stainless steel vacuum flange on the input to the downstream beam line blocked all incident FEL light from the regeneration region. Any regenerated photons would be detected in the detector system housed in the Light Tight Box (LTB) at the end of the regeneration region. (The experimental setup, as described in [27] included magnetic fields in the generation and regeneration regions. However that is not relevant here since these present results for paraphoton generation and photon regeneration are independent of magnetic field.)

The LTB was an aluminum case painted on both inner and outer surfaces with black paint, and housed inside a second box of black tape-covered aluminum foil. Inside the Light Tight Box, the photon beam passed a Newport KPX082AR16 50.2 mm lens which served to focus the photon beam to the desired accuracy onto a CCD array; the array was situated five cm downstream of the lens. The camera system was a Princeton Instrument Spec-10:400BR. It consisted of a back-illuminated CCD with a $1,340 \times 400$ pixels imaging area (a single pixel is $20\mu \times 20\mu$ in area). Data were recorded to disk using a PC.

The data acquisition system featured onboard grouping (binning) of pixels, where groups of adjacent pixels could be summed before readout to decrease noise. The detection system also consisted of a light emitting diode (LED) and a convex lens used to provide a beam spot on the CCD; this served as a reference spot on the CCD. The calibration of the CCD at the wavelength of FEL light used and at the minimum temperature (-120$^o$ C) was performed [28]. Note that any regenerated photons have the same properties as the original photons and can be focused to a small spot at the detector. Pointing stability (the direction of the laser beam relative to the central axis of the beam line) was monitored continuously during each data run. The beam was focused onto the pixel array during experimental setup. It was demonstrated in the experiment that the FEL beam could be focused to a spot size less than the diameter of a single pixel. The positions of the beam at TM2 and TM3 were monitored continuously during the data runs by cameras and Spiricon LBA-PC software. It was determined that the beam wandered by at most one cm over the two meter long beam line. This corresponds to less than 30μ of displacement at the CCD array (which is 5 cm from the focusing lens). Thus, the signal region for the pixel array was taken to be a $3 \times 3$ pixel area at the lens focus. Tests performed subsequent to the data runs confirmed that the beam focus on the signal region wandered by at most one pixel vertically and horizontally; the $3 \times 3$ pixel area defined as the signal region did not change during the data runs.

Background contributions to the signal region were studied extensively in the LIPSS setup. Data were collected with the FEL on, with and without lasing, with the CCD camera shutter open and closed in each case. Stray light from fluorescence in gas in the vacuum pipe due to cosmic rays (CR's) was shown to be negligible since the experiment was run with $10^{-6}$ Torr vacuum in the beam pipes. Stray light from all sources was shown to be less than one count per pixel per hour during the experiment. The readout noise was shown to be $2.5 \pm 0.2$ counts per pixel per readout. This contribution was minimized by collecting data for at least two hours in each run. CR's that strike the pixel array directly leave clear ionization signals in the pixels that they strike and are easily subtracted from the data. Runs that contain a CR muon hit on any pixel within an area of $100 \times 100$ pixels around the signal region were discarded. The camera system was cryogenically cooled to -120° C resulting in the lowest dark current that can be achieved under these experimental conditions, less than one single electron per pixel per hour. A check for long term drifts of the pixel thermal noise showed that this contribution was negligible over a period of several days [28].

The data were analyzed by defining a signal region where any regenerated photons would be observed, and background regions where no signal was expected. Light from a green (0.5435 microns) laser placed upstream of TM1 was focused onto the CCD array through the focusing lens shown in Fig. 2. Then, the FEL was placed in the so-called alignment mode where the laser average power was reduced by several orders of magnitude (to 0.05 per cent duty factor) so as not to damage the CCD optics and aligned in precisely the same way and focused onto the array. The lower duty factor was rigorously maintained for both machine and personnel safety when in alignment mode. In both cases, it was demonstrated that the laser light was focused by the lens down to the same, single pixel. Alignment mode runs were taken before and after the data runs, and were interspersed during the data runs in order to check for long term beam motion. No such effect was observed over the running period.

The nine pixels in the signal area were binned together in software for each run. All other pixels and pixel groupings outside the signal region were used to define the background region(s). The difference between the counts in the signal region and the counts in the background region (normalized to the number of pixels in the signal region) was determined for all data runs. No excess events above background were seen in any single run, or if all runs were combined. Twenty hours of data were taken and analyzed.

The rate of regenerated photons, $r_s$, is given by

$$r_s = n_i \cdot P_{trans} \cdot \frac{\Delta \Omega}{\Omega} \cdot \varepsilon \qquad (2)$$

where $n_i$ is the FEL (incident) photon rate, $\Delta\Omega/\Omega$ is the photon collection efficiency (solid angle for detection), $\varepsilon$ is the detector quantum efficiency, and [6]

$$P_{trans} = 16\chi^4 \left[ \sin\left(\frac{\Delta k L_1}{2}\right) \sin\left(\frac{\Delta k L_2}{2}\right) \right]^2 \qquad (3)$$

is the probability for photon regeneration from paraphotons that mix with incident photons in the generation region and propagate through the wall indicated in Fig. 1; the maximum value for $P_{trans}$ in this experiment is $1.57 \times 10^{-24}$. Here $\chi$ is the mixing parameter defined in (1), $L_1$ ($L_2$) is the length of the generation (regeneration) region shown in Fig. 2. The momentum difference between the photon and the hidden-sector paraphoton is defined as

$$\Delta k = \omega - \sqrt{\omega^2 - m_{\gamma'}^2}$$
$$\approx \frac{m_{\gamma'}^2}{2\omega} \qquad \text{for} \quad m_{\gamma'} << \omega \qquad (4)$$

where $\omega$ is the laser photon energy and $m_{\gamma'}$ is the paraphoton mass.

The 95% confidence limit is obtained from the significance of the result. This is defined as $r_s / \sqrt{r_b}$ where $r_s$ is the number of events in the signal region as described above and $r_b$ is the number of events in the pixels used as background measurements, that is, those pixels that are outside of the signal region. The background events are normalized to the same CCD array area in cases where a large area is used to get high background statistics. There is no indication of an excess of events above background for any cuts applied to the data.

The results from this run can therefore be used to set the new limits on the mixing $\chi$ of photons to hypothetical hidden-sector paraphotons as shown in Fig. 3. The full curve is the new LIPSS result, compared with those from the GammeV [22] and BMV [23] collaborations. The region above the curves is ruled out in each case. This LIPSS result represents the most stringent limits to date on this mixing in a generation-regeneration experiment in this range of parameters. The limits set by the BFRT collaboration [29] are less than those presented in Fig. 3 for each case. The new LIPSS limits are approximately a factor of three better than the best previous limit.

New optics have been installed in the JLab FEL that should yield higher laser power. The LIPSS collaboration plans to continue this work with the improved optics and beamline in order to set even more stringent limits in the near future.

## Acknowledgements

The authors thank the technical staff of the Jefferson Lab Free Electron Laser Facility, especially F. Dylla, G. Neil, G. Williams, R. Walker, D. Douglas, S. Benson, K. Jordan, C. Hernandez-Garcia, and J. Gubeli, as well as M.C. Long of Hampton University for their excellent support of the experimental program. Funding from the Office of Naval Research Award N00014-06-1-1168 is gratefully acknowledged.

**Figures**

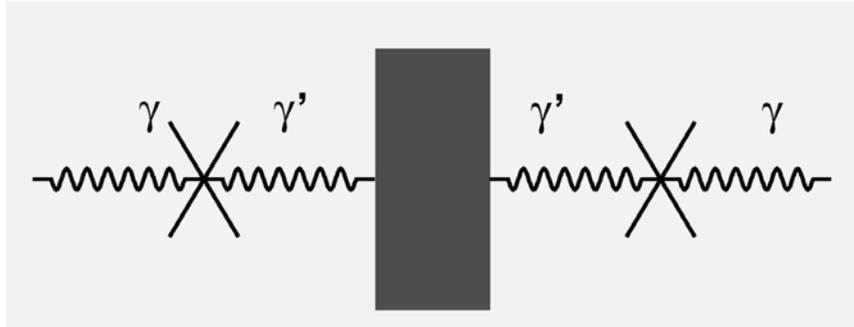

FIG. 1: Photons ($\gamma$) may convert into hidden-sector paraphotons ($\gamma'$) which proceed unimpeded through an optical barrier, reconvert back into photons downstream of the wall, and be detected in a properly executed experiment. The reconverted photons are expected to have the same properties as the original photons in this "light shining through a wall" experiment.

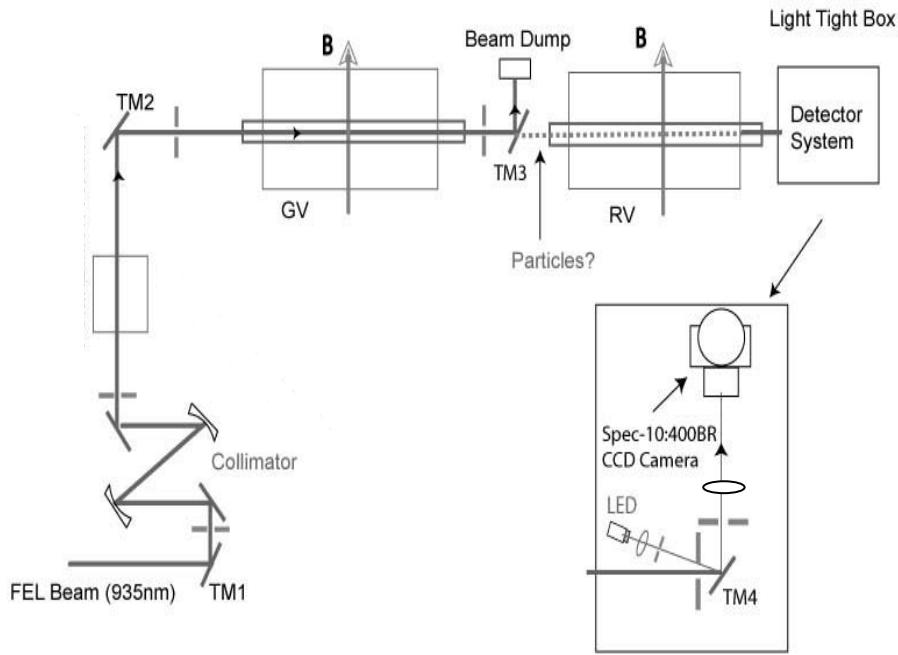

FIG. 2: The LIPSS experimental setup. Laser light from the JLab FEL is directed onto the LIPSS beamline via TM1 and a collimator. TM2 directs the properly prepared laser beam onto the generation region upstream of the beam dump at TM3. No incident photons pass through the beam dump. An identical regeneration region sits downstream of the optical barrier. Paraphotons would pass through the wall and be reconverted into photons which are then detected in the LTB.

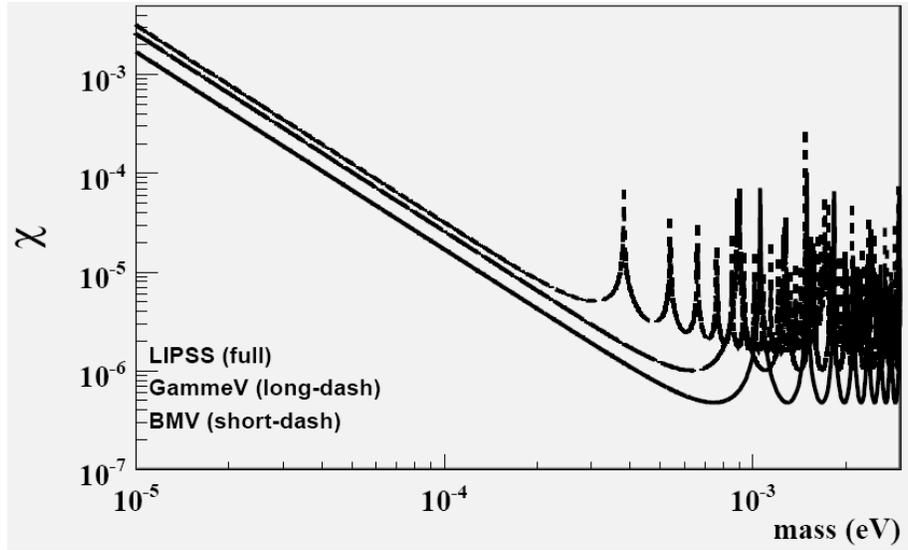

FIG. 3: A mixing parameter $\chi$ versus hidden sector paraphoton mass. Upper limits (95% confidence) set by the recent "light shining through wall" experiments [6]. The short-dashed curve is from the BMV collaboration, the long-dashed curve is from the GammeV collaboration, and the full curve is the new result from the LIPSS collaboration. The latter is approximately a factor of three better than the previous limit.